# Droplet Controlled Growth Dynamics in Plasma-Assisted Molecular Beam Epitaxy of In(Ga)N Materials

Mani Azadmand[1], Luca Barabani[1], Sergio Bietti[1], Daniel Chrastina[2], Emiliano Bonera[1], Maurizio Acciarri[1], Alexey Fedorov[3], Shiro Tsukamoto[1], Richard Nötzel[4] and Stefano Sanguinetti[1]*

1 L-NESS and Dipartimento di Scienza dei Materiali, Università di Milano-Bicocca, Milano (Italy)

2 L-NESS and Dipartimento di Fisica, Politecnico di Milano, Como (Italy)

3 L-NESS and IFN–CNR, Milano, (Italy)

4 South China Normal University, Guangzhou (China)

*Corresponding author: stefano.sanguinetti@unimib.it

## Abstract

We investigate the effect of the formation of metal droplets on the growth dynamics of InGaN by Plasma-Assisted Molecular Beam Epitaxy (PAMBE) at low temperatures ($T$ = 450°C). We find that the presence of droplets on the growth surface strongly affects the adatom incorporation dynamics, making the growth rate a decreasing function of the overall metal flux impinging on the surface as soon as the metal dose exceeds the critical amount required for the nucleation of droplets. We explain this phenomenon via a model that takes into account droplet effects on the incorporation of metal adatoms into the crystal. A relevant role is played by the vapor-liquid-solid growth mode that takes place under the droplets due to nitrogen molecules directly impinging on the droplets.

## Introduction

Among compound semiconductors, InGaN has very unique properties such as high near band edge absorption, high carrier mobility, surface electron accumulation, and superior radiation resistance [1,2]. But what puts this material in high demand for many industrial applications is its wide tunability of the band gap which spans from 0.7 to 3.4 eV, depending on the In composition. This large energy range covers almost the whole solar spectrum thus making InGaN the optimal material for solar applications [3]. InGaN characteristics are desirable for many applications such as light sources, light detectors, solar cells, photo electrochemical water splitting and electrochemical biosensors [4–6]. Moreover, growth of low to high indium composition InGaN directly on Si has recently been demonstrated [7] for cost reduction and integration with Si technology. $In_{0.5}Ga_{0.5}N$, whose bandgap lies in the near-infrared spectral region, is the most critical composition for the growth of high-quality epitaxial layers due to the strong tendency of phase separation in InGaN materials [2].
InGaN growth is hindered by the lattice mismatch and the different thermal stabilities of the two bond types present in the material: In-N and Ga-N. The lattice mismatch leads to a miscibility gap which can cause fluctuations of the In content in the epilayer [8,9].

The different binding energies of In-N and Ga-N bonds are reflected in the different decomposition temperatures of InN (630°C) and GaN (850°C) [10]. Consequently, a reduction of In incorporation in the epilayer occurs not only due to the re-evaporation of physisorbed surface adatoms but also due to the thermal decomposition of In-N bonds. Low growth temperatures have been used to avoid InN decomposition and In desorption, thus allowing the growth of high-In-composition InGaN layers by plasma-assisted molecular beam epitaxy (PAMBE) [11].
So far, many works have been devoted to the study phenomena such as phase separation [12], surface reconstruction during transition between metal-rich and N-rich regions [13], and the other growth behaviors of InGaN grown mostly at high (normal) temperatures [14–19]. There is still a lack of knowledge about the growth dynamics of InGaN at low temperature (450°C).

In this study we demonstrate the relevant role of droplets in determining the InGaN growth dynamics, by analyzing the adatoms and the kinetics of incorporation in the presence of droplets. We propose a theoretical approach that models the observed phenomena highlighting the role of the droplet as a sink for the metal adatoms and the role of the Vapor Liquid Solid (VLS) growth mode that takes place under the droplets in the InGaN epilayer growth.

## Experimental:

The InGaN thin-films were grown by molecular beam epitaxy (MBE) equipped with a radio frequency (RF) plasma source for nitrogen, on Si (111) substrates. The native silicon oxide was removed from the surface by a thermal annealing at 850°C for 30 min in vacuum. Prior to the growth, the substrates were exposed to the nitrogen plasma with a flux of 0.9 sccm (standard cubic centimeter per minute) and RF power of 360 W at 800°C. This procedure has been established to grow high quality GaN and InGaN epilayers on Si [4,20]. Subsequently the temperature of the sample was reduced to 450°C (growth temperature) and the InGaN thin film was grown with equal Ga and In fluxes (see table 1). The same N flux and RF power was maintained as during the nitridation step.
After growth, the samples were rapidly cooled down to room temperature and taken out of the MBE chamber for further characterization.
X-ray diffraction (XRD) was employed to examine the sample composition. The surface morphology was investigated by both optical microscopy as well as scanning electron microscopy (SEM) using a secondary electron detector. The layer thickness, and in turn the growth rate, was determined via cross-section SEM images of the samples.

## Results and Discussion

Figs. 1(a)-1(e) show the optical microscopy and SEM images taken from samples with different metal fluxes. In slightly N-rich conditions (excess of active N atoms compared to metal atoms

reaching the surface), the surface of the epilayer is compact and free of metal droplets. The growth rate proportionally increases with the total metal flux. XRD measurements of the samples grown in the intermediate regime, confirm 45% In in the final composition, which is very close to our target ($In_{0.5}Ga_{0.5}N$). The droplets start to appear as the ratio of metal flux vs nitrogen flux exceeds the equilibrium point, which in our experiments is $F = 0.98 \times 10^{14}$ atoms $cm^{-2}$ $s^{-1}$. The creation of metal droplets has been reported also in case of GaN grown at low temperature [21] and in general in the growth of InGaN in the metal rich phase [22].

As the droplets start to appear on the surface, we observe a marked decrease in growth rate (see Fig. 2). In addition, it is found that in metal-rich conditions, by increasing the time of growth (at the same metal/N ratio) the growth rate increases (Fig. 4b).

The creation of Ga droplets on the growth surface as a function of temperature and Ga flux has been observed in the case of GaN [23] and we observe the same behavior in our work in the case of InGaN. Here we have observed that the growth rate depends strongly on the overall metal flux and it tends to be reduced as soon as droplets are formed on the surface (Fig. 2).

At low metal flux the growth rate is proportional to the impinging flux (green shaded area in Fig. 2). These conditions correspond to N-rich conditions and the growth rate is controlled by the metal flux. The linear increase with the metal flux continues up to point where the growth rate reaches maximum value. When this point is reached, by increasing the metal flux the growth rate experiences a noticeable and continuous drop. Such a phenomenon occurs in the metal-rich growth regime and is not restricted our InGaN growth procedure, but has also been observed in the case of GaN [21]. Especially at low substrate temperatures, the excess of Ga atoms sticks on the surface and forms droplets which block GaN growth underneath. This has been shown to be detrimental to material properties and a major drawback for device applications [24] although there was no clear explanation reported for these phenomena. At higher substrate temperatures, an intermediate Ga-rich regime is observed, leading to smooth droplet-free surfaces [25].

To understand the observed behavior, we carefully take into account the metal adatom dynamics during the growth in the transition between the N-rich and the metal-rich regimes. Among all growth parameters, the III/N ratio is one of the key points to improve the surface structure and morphology of the grown film. Based on this ratio, two different morphologies have been observed. One is the compact structure which is achieved in metal-rich and lightly N-rich conditions and the other is the so-called nano-columnar structure which will appear in case of N-rich growth [26]. Nitride semiconductor films grown with even a slight excess of N during growth (N-stable conditions) display a rough surface morphology with a columnar structure initiated by the formation of stacking faults [27]. This can be explained considering the very different diffusivity for Ga and N adatoms on the surface. While Ga is very mobile at typical growth temperatures, the diffusion of N is slower by orders of magnitude. The presence of excess N strongly increases the Ga diffusion barrier. It has been calculated that the diffusion barrier for Ga adatoms on N-rich surfaces is as high as 1.8 eV, whereas it is only 0.4 eV when the growth is carried out on a Ga-saturated surface. Thus, N-rich growth leads to very low adatom mobility and to an undesired kinetically induced roughening of the surface [28]. Thus, GaN growth by PAMBE is commonly carried out under metal-rich conditions. But it must not be forgotten that in metal-rich regimes where there is the accumulation of metal atoms and creation of droplets on the surface, these very mobile metal adatoms on the surface can join metal droplets instead of participating in the crystal growth.

We observe that the decrease in the growth rate is always accompanied by the presence of metal droplets on the sample surface after the growth. The growth rate decreases more and more with increase of the metal flux and therefore the amount and volume of droplets, though the N flux is unchanged. This is in marked contrast to the common assumption that the growth rate under metal rich conditions is determined by the N flux. Hence, the common N flux calibration procedure under metal-rich conditions has to be reconsidered. The droplets thus should play a fundamental role in the change of metal adatom incorporation dynamics. As soon as the number of metal atoms exceeds the critical density for droplet formation, the metal atoms start to form droplets and accumulate in them. Then, the presence of droplets on the surface establishes a depletion channel for the metal adatom density on the growth surface, as metal adatoms can be efficiently captured by the droplets. This depletion channel is in competition with the N-driven metal adatom incorporation into the InGaN crystal. This leads, on one side, to a reduction of the metal adatom incorporation rate into the crystal in the regions not covered by the droplets. On the other side, the presence of liquid metal droplets on the surface should allow for the VLS growth of InGaN material under the droplets themselves, via direct incorporation of nitrogen. In summary, three possible processes are available to the metal adatom on the growing InGaN surface in the presence of droplets (see Fig.3a):

1) Desorption
2) Incorporation of metal adatom into the crystal by binding to a N active site on the droplet-free surface
3) Metal adatom attachment to a droplet

In addition, the metal adatoms incorporated in the droplets can be incorporated in the crystal due to VLS mechanism under the droplet by direct N flux into the droplet. It is possible to model the combined effect of the three processes as a set of rate equations for the metal adatom density *n* and VLS growth rate (see **Error! Reference source not found.**b). The first equation describes the kinetics of the metal adatom density *n*, determined by the combined effect of external metal flux, incorporation in to the crystal and attachment to the droplets:

$$\frac{\boldsymbol{dn}}{\boldsymbol{dt}} = \boldsymbol{F} - \boldsymbol{E} - \boldsymbol{n\mu}(\boldsymbol{\Phi}/\mathbf{F}) - \boldsymbol{n\sigma\varrho}(\boldsymbol{F}) \qquad (1)$$

Where $F$ is the total (In plus Ga) metal flux, $E$ is the desorption flux, $\mu(\Phi/F)$ the incorporation probability into the crystal on the droplet free surface, which depends on the ratio between active N flux $\Phi$ and the metal flux $F$. $\mu(\Phi/F)$ is proportional to the probability, for a metal adatom, to find an active N site for binding. $\rho(F)$ the droplet density and σ the droplet capture cross section. At low temperature in MBE conditions the metal desorption flux is close to zero ($E$ = 0). Therefore in the steady state condition, Eq. (1) leads to the solution for $n$

$$n = F\,\frac{1}{\mu(\Phi/F) + \sigma\rho} \qquad (2)$$

And consequently, the bulk growth rate per unit surface area (process 2) $R_S$ that takes place in the regions of the sample where there are no droplets is:

$$R_S(F,\Phi) = n\mu(\Phi/F) = F\frac{1}{1+\frac{\sigma\varrho_o F^p}{\mu(\Phi/F)}} \qquad (3)$$

Here we explicitly used the expected dependence of the droplet density on the metal flux $\rho(F)=\rho_0 F^p$ [29]. $p$ is the critical exponent for droplet nucleation, whose value is determined by the physics of the nucleation process. As droplet formation happens in the presence of a leak channel for the adatoms, that is the bulk incorporation, we expect to be in the regime of incomplete condensation. In this regime $p > 1$ values are expected. It is worth mentioning that in the N-rich region, where the density of droplets is equal to zero (green region in Fig. 2), Eq. (3) reduces to:

$$R_S(F,\Phi) = F$$

In N-rich conditions the predicted InGaN growth rate linearly follows the metal flux, as observed. When the growth turns to metal-rich conditions, as soon as the adatom density $n$ overcomes the critical density for droplet formation, the second depletion channel opens for the metal adatoms (in addition to the incorporation), thus leading to a rapid depletion of the adatom density $n$. The depletion of $n$ leads a reduction of $R_S$ in the area between the droplets. The effect becomes larger as the metal flux increases, as the droplet density, and thus in turn the metal adatom capture probability by the droplets, increases with $F$. The transition between the two growth modes takes place around $F_T = 1 \times 10^{14}$ atoms cm$^{-2}$s$^{-1}$.

However, under the droplets, VLS growth can take place, due to the incorporation of nitrogen molecules into the droplets by direct impingement [30]. The growth rate per unit area is $R_U(\Phi)$ and depends on the nitrogen flux $\Phi$, the incorporation probability of N into the droplet and the growth rate at the liquid-solid interface. The total growth rate is therefore the sum of the two contributions, that is the growth rate $R_S$ in the droplet free areas and $R_U$ under the droplets, whose relevance is given by the surface area covered by the droplets (see Fig.4a). The total growth rate per unit area is then:

$$R_T(F,\Phi,\tau) = (1-\chi)R_S + \chi R_U = R_S + \chi(R_U - R_S) \qquad (4)$$

where $\chi$ is the relative area covered by the droplets. The surface area covered by the droplets depends on the total amount of metal stored in the droplet ensemble and on the droplet density $\rho$. As long as the droplets do not touch ($\chi \ll 1$), the dependence of $\chi$ on the metal flux $F$, the deposition time $\tau$ and the droplet density $\rho$ is:

$$\chi(\tau) = \gamma\left[\frac{(F-R_S)\tau}{\rho}\right]^{2/3} \qquad \rho = \gamma\varrho_0^{1/3}\left(1-\frac{1}{1+\frac{\sigma\varrho_o F^p}{\mu\left(\frac{\Phi}{F}\right)}}\right)^{2/3} F^{\frac{2+p}{3}}\tau^{2/3} \qquad (5)$$

where γ is a normalization constant which can be derived from experiments by equating the Eq. (5) predictions with the observed $\chi$ at maximum metal flux. $(F-R_S)\tau$ is the total metal quantity that is stored in the droplet ensemble after the growth time τ. It corresponds to the total metal dose deposited on the substrate minus the amount incorporated in the crystal between the droplets ($R_S$). We did not consider the decrease in the metal dose available for droplet formation due to the crystallization under the droplet as it is a small correction for $\chi \ll 1$ and it adds complexity to the model. It is worth mentioning that Eq. (5) is not valid for long growth times, when $\chi \rightarrow 1$ and the amount of growth proceeding via VLS under the droplets becomes relevant. By combining equations (3-5) we find the dependence of the total growth rate, at time τ, per unit area to be given by the relation

$$R_T = F\,\frac{1}{1+\frac{\sigma\varrho_o F^p}{\mu(\Phi/\mathrm{F})}} + \gamma\left[\varrho_0^{1/3}\left(R_U(\Phi) - F\,\frac{1}{1+\frac{\sigma\varrho_o F^p}{\mu(\Phi/F)}}\right)\left(1-\frac{1}{1+\frac{\sigma\varrho_o F^p}{\mu(\Phi/F)}}\right)^{2/3} F^{\frac{2+p}{3}}\right]\tau^{2/3} \tag{6}$$

The total growth rate $R_T$ *versus* time dependence is introduced by the change, with the growth time, of the area covered by the droplets. The exact $\tau^{2/3}$ dependence is related to the invariance of the wetting angle of the droplet with its size. In the extreme case of a surface covered by closely arranged droplets, the growth rate will be determined by the N flux only, that is $R_T \propto \Phi$. It is worth mentioning that the growth progressively shifts towards the VLS dominated mode as the growth time increases in metal-rich conditions.

From equation (6) the dependence on time of the average growth rate Γ is

$$\Gamma = R_S + \frac{3\gamma}{5}B(F,\Phi)\tau^{2/3} \tag{7}$$

where $B(F,\Phi)$ indicates the complex formula encased in the square brackets in Eq. (6). The model predictions, using Eq. (6) and (7), are reported in Fig. 2. After the initial linear dependence on $F$, the predicted curve shows a clear decrease in the growth rate. The good agreement of the model predictions based on the dependence of $R_S$ on $F$ (black curve) indicates that the main factor determining the reduction of the growth rate is the decrease in the adatom density $n$ caused by the opening of the droplet depletion channel. The critical exponent extracted from the model is $p = 1.6$. The observation of an exponent $p > 1$ is the outcome of the droplet nucleation in the regime of incomplete condensation. This is to be expected since in presence of nitrogen, a fraction of the deposited metal flux is incorporated into the crystal, and therefore cannot contribute to droplet nucleation or to the increase in volume of existing droplets. We find that the probability for a metal adatom to attach to a forming droplet is three times larger with respect to being incorporated in the growing crystal at $F_T$ ($\sigma\rho_0 = 3\mu$). This significant difference in incorporation probability leads to the observed fast decrease of the average growth velocity and its lack of recovery even at high $F$. Considering also VLS growth under the droplet (red curve in Fig. 2), $R_T$ still decreases with the increasing metal flux, although a sizeable decrease in the curve slope with $F$, eventually reaching a plateau at high metal fluxes, is clearly present. The VLS growth mode dominates the dynamics when the droplet coverage reaches a sizeable percentage of the surface. According to our experimental observations, the maximum coverage in our samples is $\chi_M = 0.6$ (sample K, see Table I). One of fitting parameter of the model is the growth rate per unit area under the droplets ($R_U$). We find that $R_U$ is equal, at $F_T$, to $R_S$. This suggests that the incorporation rate of N into the droplets and the InGaN surface at the growth temperature of 450 °C and $F_T$ are equivalent.

The model predicts a recovery of the growth rate as the growth time increases, with a $\tau^{2/3}$ dependence. This is due to the increase of the area covered by the droplets whose outcome is the increase in relevance of the VLS mode. To test this expected phenomenon we grew three samples, at $F$ = 2.94 × $10^{14}$ atoms $cm^{-2}s^{-1}$, with different growth times (see Fig. 4b). In this sample series, a clear dependence of the average growth time is clearly observed. The behavior of $\Gamma$ when $\tau$ increases follows the predictions of Eq. (7), thus conforming the model predictions that in the metal-rich zone the growth is increasingly dominated by the VLS mode in the regions covered by the droplets.

## Conclusion

The observed dynamics in the growth rate in the metal rich region is determined by the droplets. This happens in two ways. The first is related to the metal adatom attachment to the droplet that depletes the adatom density $n$ thus hindering the epilayer growth. This phenomenon becomes more and more relevant as the metal flux increases, as the surface is increasingly covered by droplets, thus leading to the observed reduction of the growth rate. On the other side, the presence of the droplets on the surface activates the VLS crystal growth under the droplet. This relates to direct impingement of N onto the droplet. Our findings show that this growth method is faster than the normal crystal incorporation and promotes, at long times, a recovery of the growth rate in samples with high droplet coverage.

# Tables

*Table1*

*Metal flux (In + Ga with one to one ratio), growth time, and growth rate of samples*

| Sample | Metal flux ($10^{14}$ atoms $cm^{-2}s^{-1}$) | Growth time (mins) | Metal dose ($10^{16}$ atoms $cm^{-2}$) | Growth rate (ML/s) |
|---|---|---|---|---|
| A | 0.39 | 120 | 28 | 0.0401 |
| B | 0.78 | 120 | 56 | 0.0740 |
| C | 0.98 | 120 | 71 | 0.1028 |
| D | 1.17 | 120 | 84 | 0.0900 |
| E | 1.41 | 90 | 76 | 0.0802 |
| F | 1.96 | 45 | 53 | 0.0843 |
| G | 2.94 | 45 | 79 | 0.0813 |
| H | 2.94 | 90 | 158 | 0.0874 |
| I | 2.94 | 150 | 264 | 0.1028 |
| J | 3.92 | 60 | 141 | 0.0637 |
| K | 7.83 | 90 | 423 | 0.0510 |

# Figures

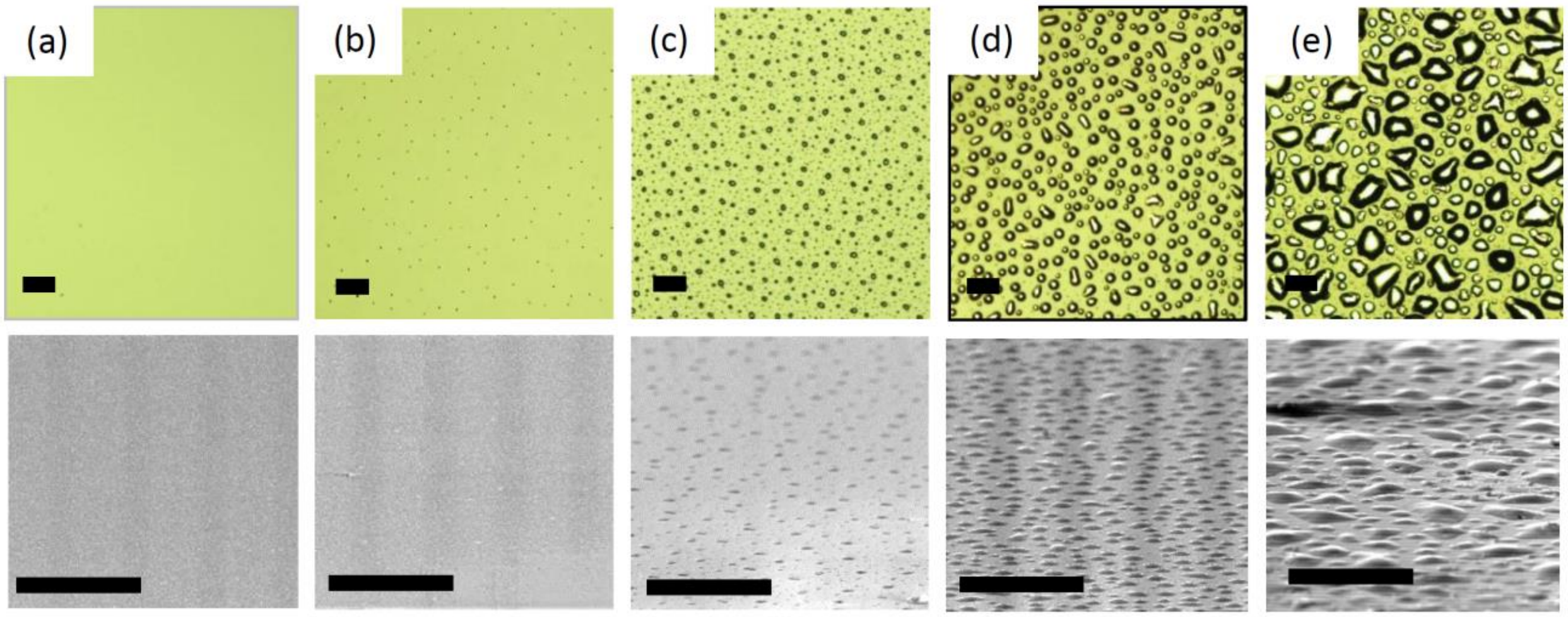

*Figure 1: (top) Optical Microscope and (bottom) related SEM images (taken at an angle of 70° respect to the sample) of InGaN samples grown under different metal fluxes, from left to right: (a) $0.39 \times 10^{14}$, (b) $0.98 \times 10^{14}$, (c) $1.41 \times 10^{14}$, (d) $3.92 \times 10^{14}$, (e) $7.83 \times 10^{14}$ atoms $cm^{-2}$ $s^{-1}$. Black line length in all the images is 100 µm.*

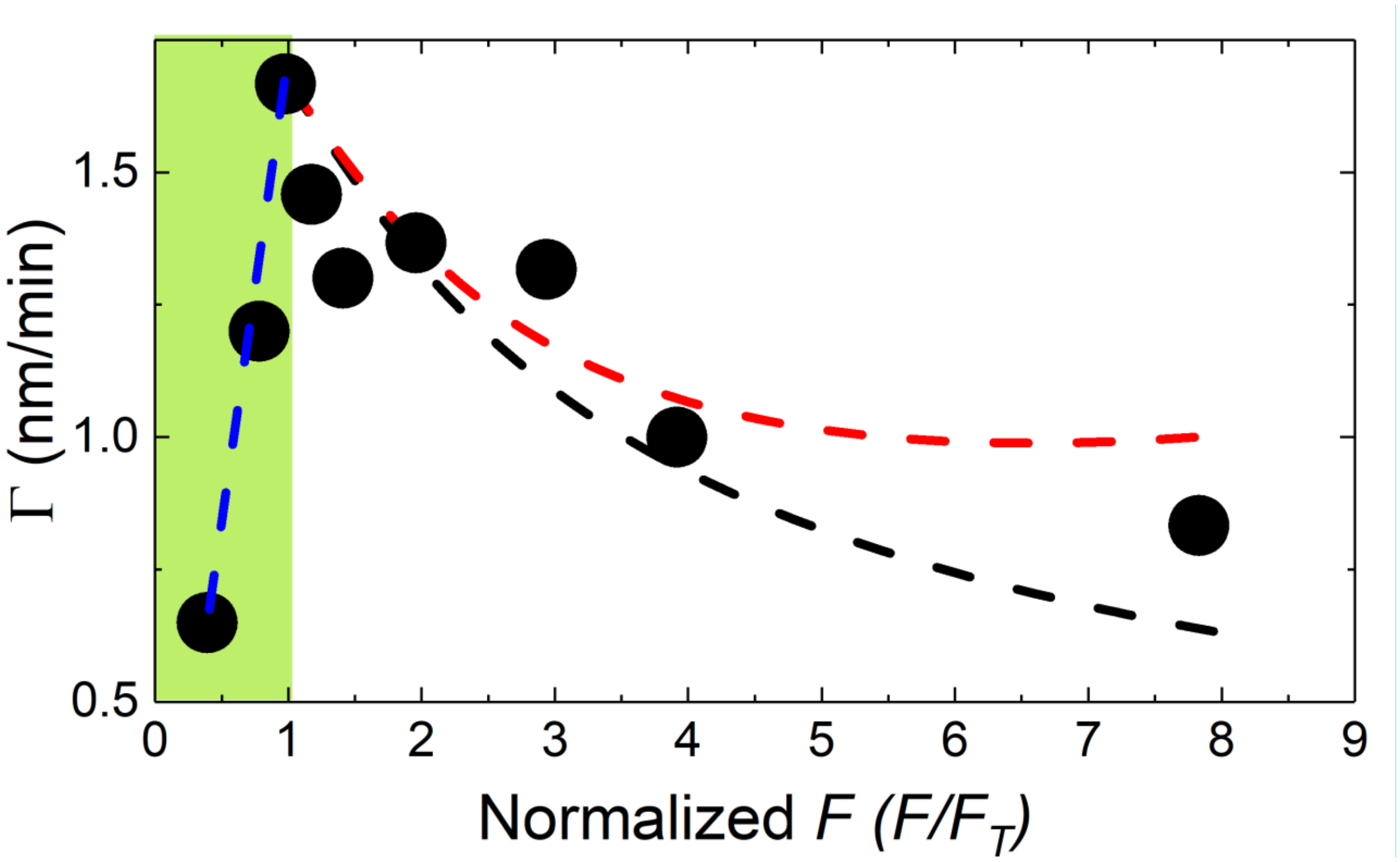

*Figure 2: Average growth rate as a function of metal flux F normalized by $F_T$. The dashed lines correspond to the model predictions: Initial linear dependence in absence of droplets (blue), Γ (red) and $R_S$ (black). Model parameters are $R_U = R_S(F_T)$, $\chi_M = 0.6$, $\Gamma_{MAX} = 1.7$ nm/min, $\sigma\rho_0/\mu = 3$, the critical exponent $p = 1.6$.*

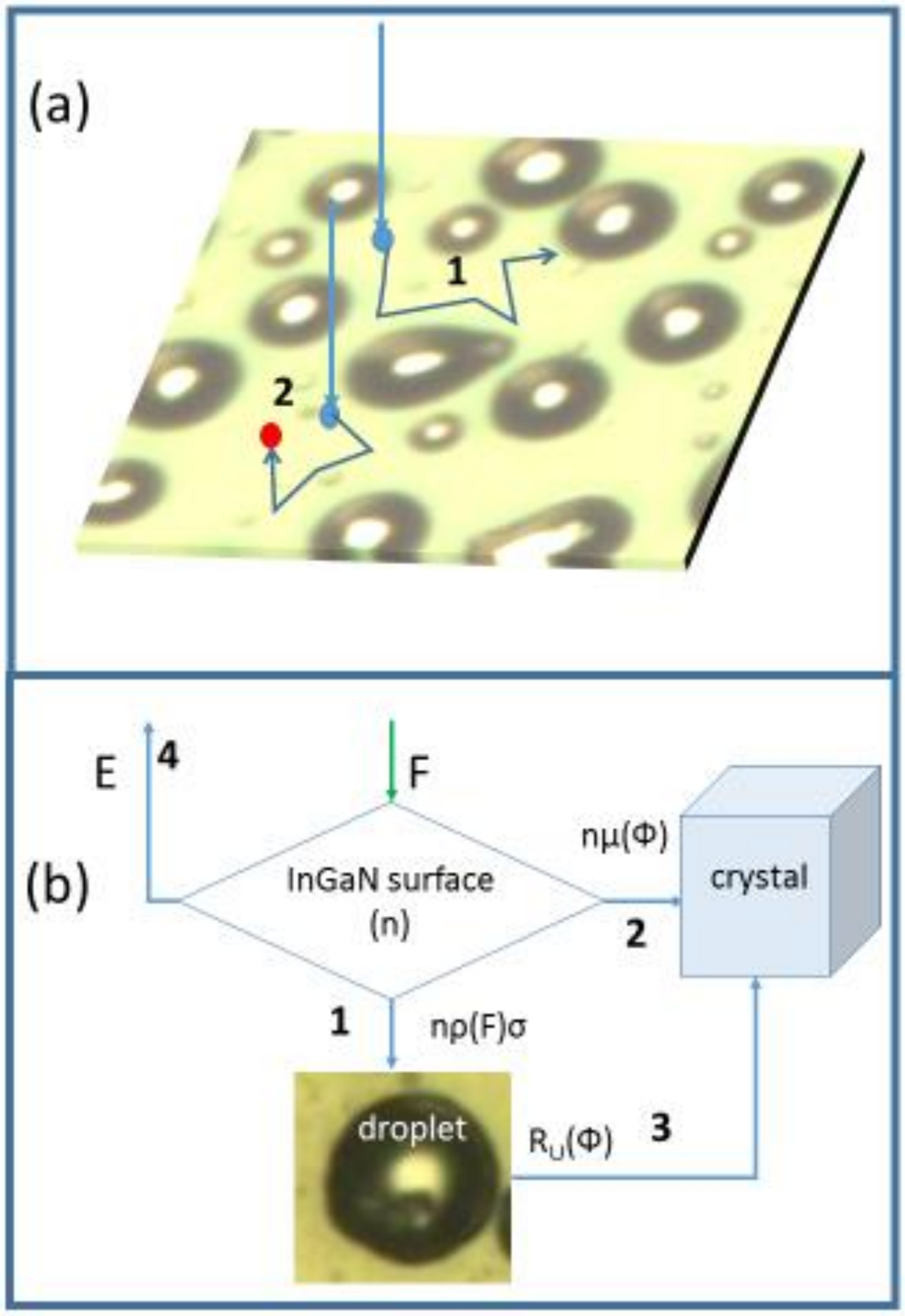

*Figure 3: (a) Available metal adatom processes. Process 1 leads to metal droplet attachment. Process 2 leads to incorporation into the growing crystal (indicated by the red dot). (b) Schematics of the metal adatom rate equation model highlighting the four available channels for the metal adatoms: 1) droplet attachment; 2) crystal incorporation; 3) the VLS crystal growth that takes place under the droplets. 4) desorption channel*

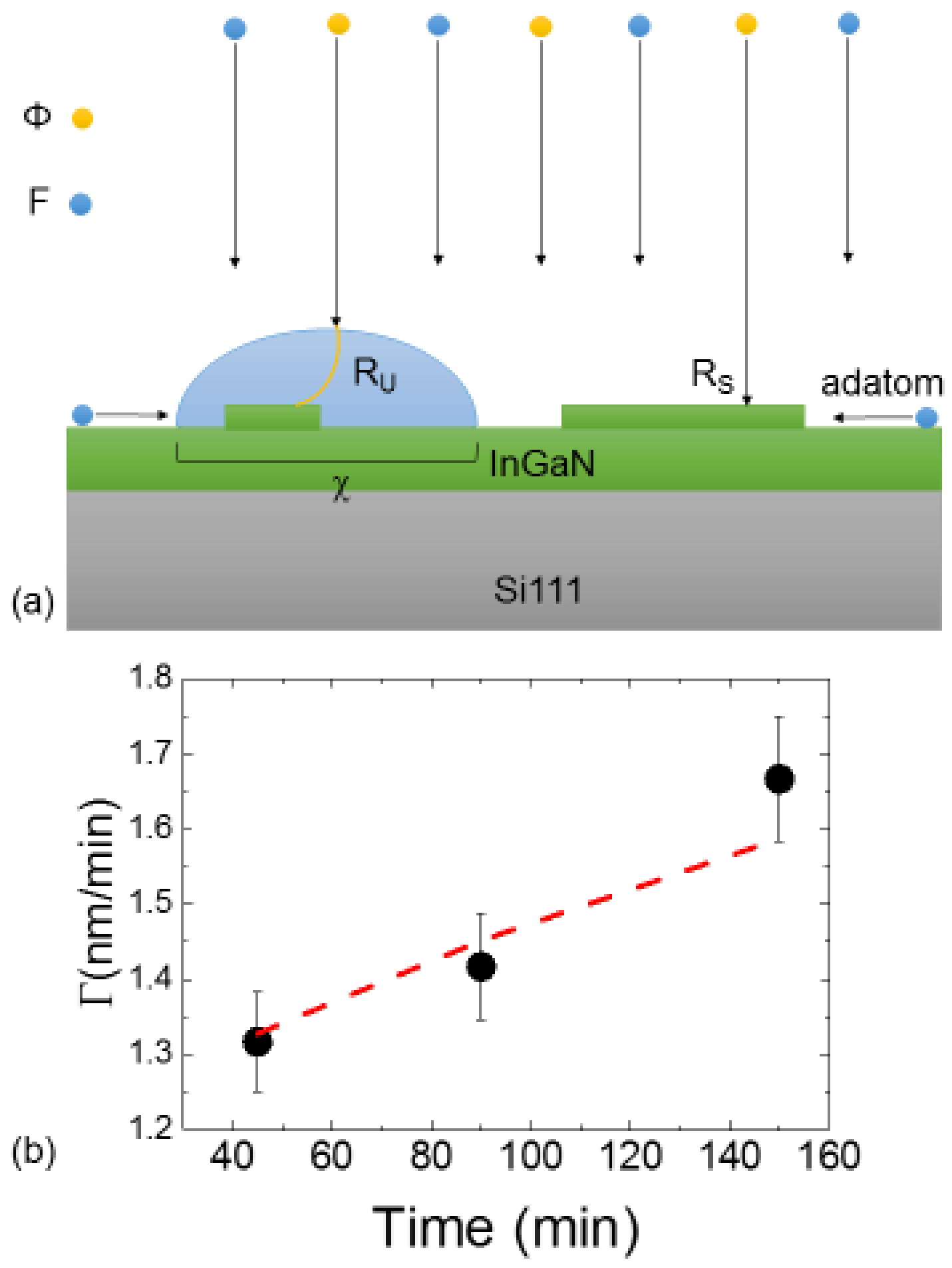

*Figure 4: (a) Schematic of crystal growth under and between the metal droplets. (b) Observed dependence of average growth rate from growth time (black dots). The dashed line represents the model predictions [Eq. (7)]. Here $R_C = 67$ nm/h and $3\gamma B/5 = 15$ nm/h$^{1/3}$. Metal flux used in the growth: $F = 2.94\times 10^{14}$ atoms cm$^{-2}$s$^{-1}$.*